\documentclass{ptephy_v1}%%%%%% to generate preprint number with ptep logo

\preprintnumber{XXXX-XXXX} %%% %%% Insert preprint number here

\usepackage{amsmath} %for dealing with mathematics,
\usepackage{amsthm} %for dealing with theorem environments,

\usepackage{bm}

\begin{document}

\title{Geometrical structure and thermal conductivity of dust aggregates formed via ballistic cluster-cluster aggregation}

%%%% To generate auto affiliation numbers please use \author{}\affil{} command

\author{Sota Arakawa}
\author{Masaki Takemoto}
\author{Taishi Nakamoto}
\affil{Department of Earth and Planetary Sciences, Tokyo Institute of Technology, 2-12-1 Ookayama, Meguro, Tokyo, 152-8551, Japan. \email{arakawa.s.ac@m.titech.ac.jp}}

%%% To include the collaborator name... Please use the command "\collaborator"
%%% For example: \collaborator{ATLAS Collaboration}

\begin{abstract}%
We herein report a theoretical study of the geometrical structure of porous dust aggregates formed via ballistic cluster-cluster aggregation (BCCA).
We calculated the gyration radius $R_{\rm gyr}$ and the graph-based geodesic radius $R_{\rm geo}$ as a function of the number of constituent particles $N$.
We found that $R_{\rm gyr} / r_{0} \sim N^{0.531 \pm 0.011}$ and $R_{\rm geo} / r_{0} \sim N^{0.710 \pm 0.013}$, where $r_{0}$ is the radius of constituent particles.
Furthermore, we defined two constants that characterize the geometrical structure of fractal aggregates: $D_{\rm f}$ and $\alpha$.
The definition of $D_{\rm f}$ and $\alpha$ are $N \sim {( R_{\rm gyr} / r_{0} )}^{D_{\rm f}}$ and ${R_{\rm geo}} / {r_{0}} \sim {\left( {R_{\rm gyr}} / {r_{0}} \right)}^{\alpha}$, respectively.
Our study revealed that $D_{\rm f} \simeq 1.88$ and $\alpha \simeq 1.34$ for the clusters of the BCCA.

In addition, we also studied the filling factor dependence of thermal conductivity of statically compressed fractal aggregates.
From this study, we reveal that the thermal conductivity of statically compressed aggregates $k$ is given by $k \sim 2 k_{\rm mat} {( r_{\rm c} / r_{0} )} \phi^{(1 + \alpha) / (3 - D_{\rm f})}$, where $k_{\rm mat}$ is the material thermal conductivity, $r_{\rm c}$ is the contact radius of constituent particles, and $\phi$ is the filling factor of dust aggregates.

\end{abstract}

\subjectindex{E21, I04, I10, J44}

\maketitle

\section{Introduction}

The study of the aggregation of small dust particles into larger aggregates is crucial for understanding the fundamental processes in astrophysics and geophysics.
For example, the growth of aerosol or haze particles in the atmosphere causes the scattering and absorption of the sunlight \cite[e.g.,][]{Sorensen2001}.
In addition to this, the aggregation of dust particles also occurs in the mineral clouds of exoplanets and hence understanding the fundamental processes involved in the formation process of dust aggregates in exoplanets is imperative to interpret the transmission spectra \cite[e.g.,][]{Ohno+2018,Ohno+2019}.
In addition, the aggregation of dust particles in the solar nebula is the first step towards the formation of the planets \cite[e.g.,][]{Adachi+1976,Dominik+1997,Wada+2008,Tanaka+2012}, and the density evolution of dust aggregates is considered to be the key to understanding the evolution from nm- or $\mu$m-sized dust grains to km-sized small bodies \cite[e.g.,][]{Okuzumi+2012,Arakawa+2016,Tsukamoto+2017,Tatsuuma+2018}.
The resulting aggregates frequently have a complex and fractal structure with an extremely low filling factor \cite[e.g.,][]{Meakin1988,Meakin1991,Blum+2008}.
Therefore, understanding the geometrical structure of these fractal aggregates and its influence on the physical properties such as the thermal conductivity and the compressive strength is of immense current interest.

For porous dust aggregates composed of micron-sized ${\rm Si}{\rm O}_{2}$ glass grains, the thermal conductivity is obtained by several experimental studies \cite[e.g.,][]{Krause+2011,Sakatani+2017}, and it is empirically known that the thermal conductivity is approximately proportional to the square of the filling factor \cite{Kobayashi+2013,Arakawa+2017,Arakawa+2019}.
However, the theoretical explanation of the dependence of thermal conductivity on the filling factor is still lacking.

In contrast to the thermal conductivity, the filling factor dependence of the tensile strength of porous dust aggregates is well understood \cite{Tatsuuma+2019}.
The tensile strength of porous dust aggregates is evaluated from the fractal structure of dust aggregates and the maximum force required to separate two sticking particles.
The compression strength would also be evaluated from the fractal structure of dust aggregates and the rolling energy needed to rotate a constituent particle around its connecting points \cite{Kataoka+2013a}.

In this paper, we describe the geometrical structure of porous dust aggregates formed in astrophysical environments.
We present the calculation of the gyration radius (which is defined in Section \ref{sec2.1}) and the graph-based geodesic radius (which is defined in Section \ref{sec2.2}) of porous dust aggregates.
Subsequently, we present the interpretations of the filling factor dependence of thermal conductivity from the geometrical structure, and we also confirm the validity of our theoretical understanding by comparing it with the result of direct numerical calculations in Section \ref{sec3}.
Finally, we also present the modified interpretation of the filling factor dependence of compression strength and the average coordination number of dust aggregates in Section \ref{sec4}.

\section{Ballistic cluster-cluster aggregation}

\begin{figure}
\centering
\includegraphics[width=3.0in]{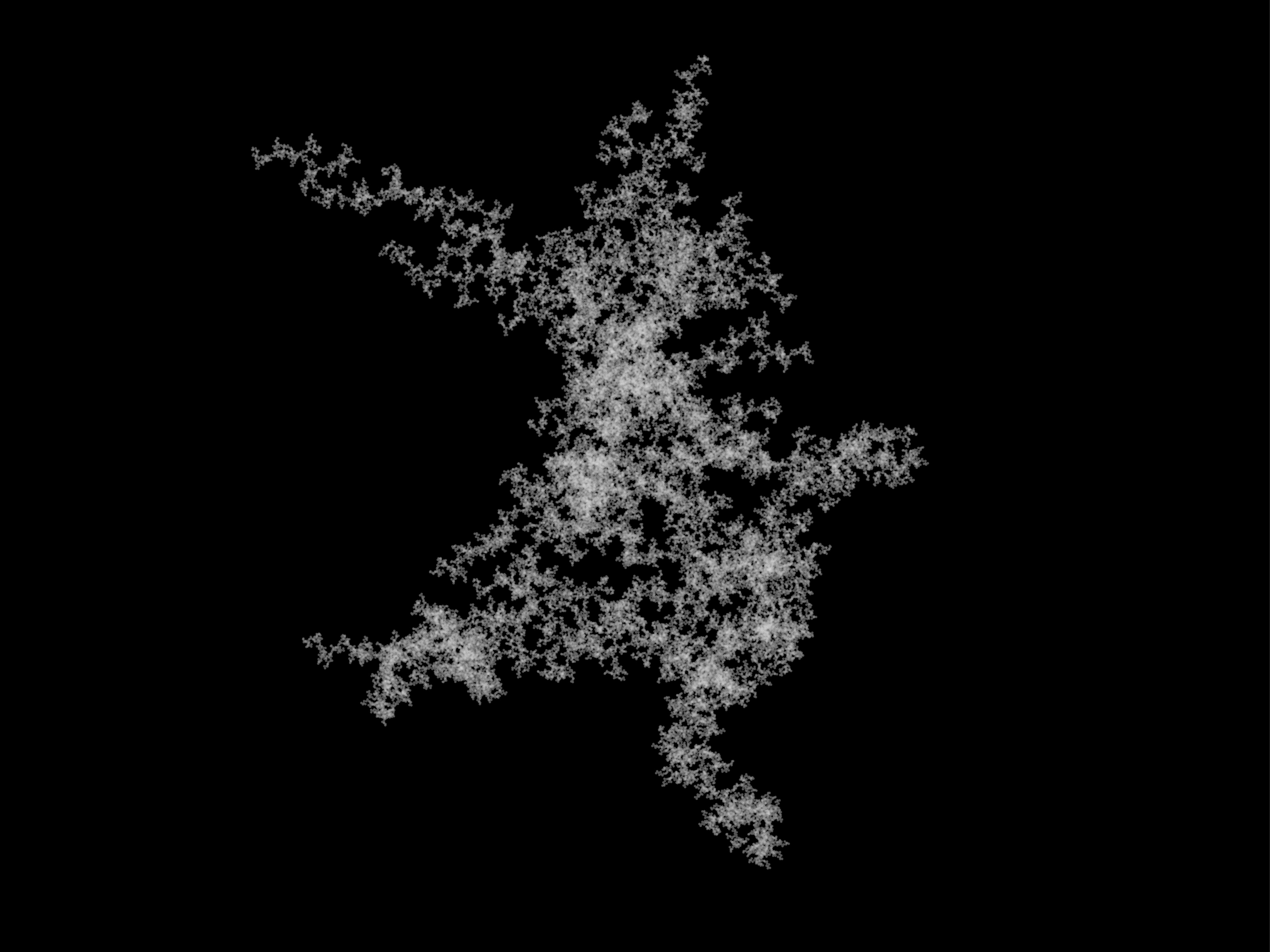}%
\caption{
Projection of a three-dimensional dust aggregate formed via ballistic cluster-cluster aggregation (BCCA).
The number of constituent particles is $N = 2^{16} = 65536$.
}
\label{fig1}
\end{figure}

In the early stage of dust growth in astrophysical environments such as protoplanetary disks and circumplanetary disks, the collision velocity is sufficiently low to avoid collisional compaction and collisions between similar-sized dust aggregates are dominant \cite[e.g.,][]{Wurm+1998,Kempf+1999}.
Therefore, the shape of dust aggregates in astrophysical environments is expected to resemble the clusters of ballistic cluster-cluster aggregation (BCCA).
The BCCA clusters are formed by the sticking of two equal sized BCCA clusters with no restructuring (see Figure 3(a) of Okuzumi et al.\ \cite{Okuzumi+2009}).
We prepare BCCA clusters in the following procedure: (1) prepare an aggregate composed of $2^{i}$ particles (initially $i = 0$), (2) copy this aggregate and change the orientation of the copy randomly, (3) by ballistic sticking of these two aggregates with a randomly chosen offset, make a new aggregate composed of $2^{i + 1}$ particles, (4) continue the procedure (1)--(3).
As shown in Figure \ref{fig1}, a BCCA cluster has a highly porous structure.
In this study, we assumed that all constituent particles (hereinafter referred to as ``monomers'') are spherical and have the same radius $r_{0}$.

\subsection{Gyration radius}
\label{sec2.1}

For the quantitative analysis of the structure of porous aggregates, we need to define a typical cluster radius. 
Here, we use the gyration radius $R_{\rm gyr}$, which is customary in aggregation studies \cite[e.g.,][]{Kempf+1999,Okuzumi+2009,Mukai+1992} defined by
\begin{equation}
R_{\rm gyr} \equiv {\left( \frac{1}{2 N^{2}} \sum_{i}^{N} \sum_{j}^{N} {\left( {\bm x}_{i} - {\bm x}_{j} \right)}^{2} \right)}^{1/2} \equiv {\left( \frac{1}{N} \sum_{i}^{N} {\left( {\bm x}_{i} - {\bm x}_{\rm o} \right)}^{2} \right)}^{1/2},
\end{equation}
where $N$ is the number of constituent monomers, ${\bm x}_{i}$ is the coordinate of the $i$-th monomer, and ${\bm x}_{\rm o}$ is the coordinate of the center of mass.

\begin{figure}
\centering
\includegraphics[width=3.0in]{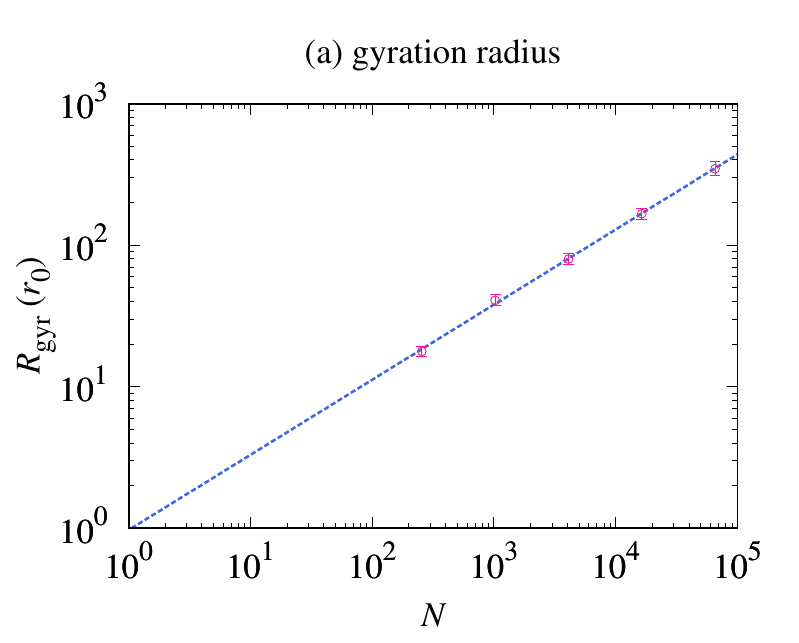}%
\includegraphics[width=3.0in]{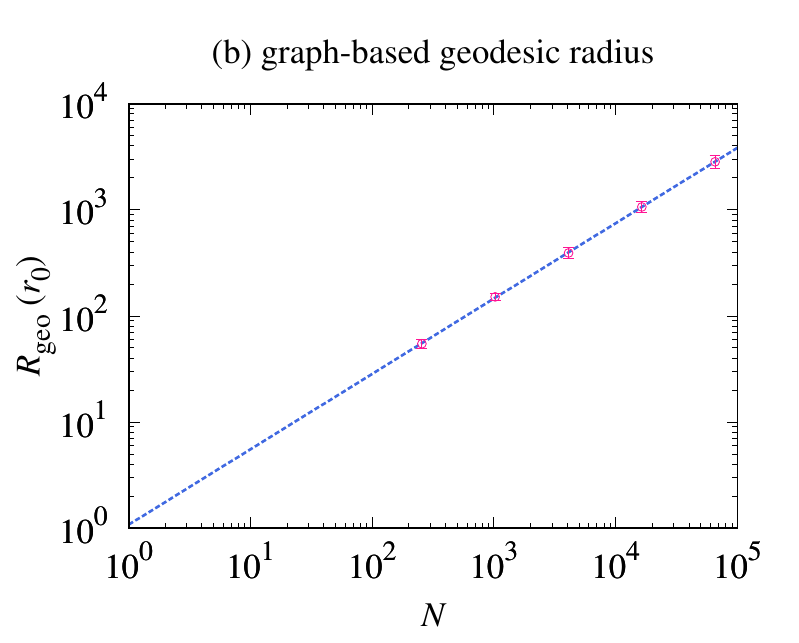}%
\caption{
(a) Fitting of the gyration radius of BCCA clusters $R_{\rm gyr}$ as a function of the number of monomers $N$.
(b) Fitting of the graph-based geodesic radius of BCCA clusters $R_{\rm geo}$ as a function of $N$.
The circles represent the averaged data with vertical error bars of twice the standard error.
The dashed line is the best-fit obtained from the weighted least-squares method.
}
\label{fig2}
\end{figure}

We carried out 20 growth sequences of $N$-body simulations of BCCA as previous studies \cite{Okuzumi+2009,Mukai+1992}. 
Here we show the geometric mean of the gyration radius $R_{\rm gyr}$ as the function of the number of monomers $N$ in Figure \ref{fig2}(a).
We found that the gyration radius $R_{\rm gyr}$ is given by
\begin{equation}
\log_{10} \frac{R_{\rm gyr}}{r_{0}} = {\left( 0.531 \pm 0.011 \right)} \log_{10} N + {\left( -0.012 \pm 0.006 \right)},
\end{equation}
and given uncertainties are the standard errors.
The structure of BCCA clusters is therefore described in terms of the fractal dimension $D_{\rm f}$, which is defined as
\begin{equation}
N \sim {\left( \frac{R_{\rm gyr}}{r_{0}} \right)}^{D_{\rm f}}.
\end{equation}
Our numerical data shows that $D_{\rm f}$ is
\begin{equation}
D_{\rm f} \simeq \frac{1}{0.531} \simeq 1.88,
\end{equation}
or $1.85 < D_{\rm f} < 1.92$ when we take uncertainty into consideration.
This result is consistent with previous studies \cite{Okuzumi+2009,Mukai+1992,Tazaki+2016}.

The effective volume of BCCA clusters $V$ is evaluated as $V \sim {4 \pi {R_{\rm gyr}}^{3}} / 3$ and the volume of monomers is $V_{0} = {4 \pi {r_{0}}^{3}} / 3$.
The filling factor of BCCA clusters $\phi$ is given by
\begin{equation}
\phi = \frac{N V_{0}}{V} \sim \frac{N}{{\left( R_{\rm gyr} / r_{0} \right)}^{3}} \sim N^{1 - 3 / D_{\rm f}}.
\end{equation}
Therefore, we can calculate the filling factor of BCCA clusters from the number of monomers.

\subsection{Graph-based geodesic radius}
\label{sec2.2}

\begin{figure}
\centering
\includegraphics[width=2.5in]{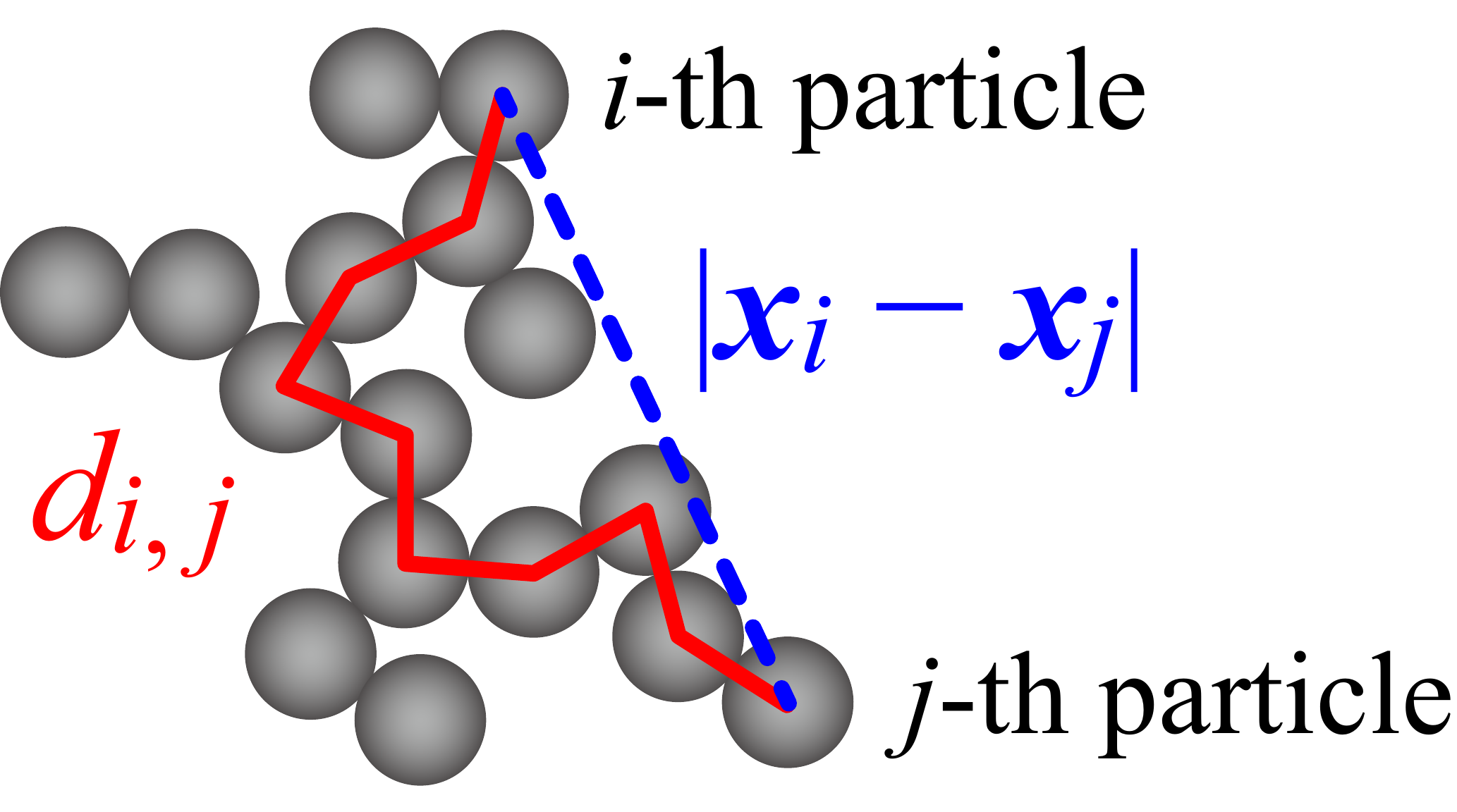}%
\caption{
Schematic description of the distance between $i$-th and $j$-th particles, $| {\bm x}_{i} - {\bm x}_{j} |$, and the graph geodesic between $i$-th and $j$-th particles, $d_{i,j}$.
It is clear that the graph geodesic $d_{i,j}$ is larger than the distance $| {\bm x}_{i} - {\bm x}_{j} |$.
}
\label{fig3}
\end{figure}

Granular materials and dust aggregates transmit compressive stresses via a network of force chains \cite[e.g.,][]{Liu+1995}.
Further, the chains of monomers also conduct heat \cite[e.g.,][]{Sirono2014}.
Therefore, understanding the chain structure within dust aggregates is essential, especially for the study of the mechanical and heat transfer properties.
Here, we introduce {\it the graph geodesic} and {\it the graph-based geodesic radius}.
Figure \ref{fig3} schematically illustrates a BCCA cluster.
The distance between $i$-th and $j$-th particles is $| {\bm x}_{i} - {\bm x}_{j} |$, and we define the graph geodesic between $i$-th and $j$-th particles as $d_{i,j}$ in Figure \ref{fig3}.
Considering the graph structure of BCCA cluster, which is a tree (i.e., a connected acyclic graph), we can uniquely determine the graph geodesic $d_{i,j}$.

The typical length of the chain of monomers is obtained using the same method as the definition of $R_{\rm gyr}$.
We define the graph-based geodesic radius $R_{\rm geo}$ as
\begin{equation}
R_{\rm geo} \equiv {\left( \frac{1}{2 N^{2}} \sum_{i}^{N} \sum_{j}^{N} {d_{i,j}}^{2} \right)}^{1/2}.
\end{equation}
It is clear that ${d_{i,j}}^{2} \geq {\left( {\bm x}_{i} - {\bm x}_{j} \right)}^{2}$ and then $R_{\rm geo} \geq R_{\rm gyr}$ by definition.
Here we show the geometric mean of the graph-based geodesic radius $R_{\rm geo}$ as a function of the number of monomers $N$ over 20 runs in Figure \ref{fig2}(b).
We found that the graph-based geodesic radius $R_{\rm geo}$ is given by
\begin{equation}
\log_{10} \frac{R_{\rm geo}}{r_{0}} = {\left( 0.710 \pm 0.013 \right)} \log_{10} N + {\left( 0.034 \pm 0.007 \right)},
\end{equation}
and given uncertainties are the standard errors.

Here, we consider the ratio of $R_{\rm geo}$ and $R_{\rm gyr}$.
The ratio of $R_{\rm geo}$ and $R_{\rm gyr}$ is given by
\begin{equation}
\frac{R_{\rm geo}}{r_{0}} \sim {\left( \frac{R_{\rm gyr}}{r_{0}} \right)}^{\alpha},
\end{equation}
where $\alpha$ is the dimensionless constant and the constant $\alpha$ must depend on the aggregation process of clusters.
For BCCA clusters, we found that
\begin{equation}
\alpha \simeq \frac{0.710}{0.531} \simeq 1.34,
\end{equation}
or $1.29 < \alpha < 1.39$ when we take uncertainty into consideration.

\subsection{Bifractality of statically compressed BCCA clusters}

In the early stage of dust growth, the fractal dimension of dust aggregates is $D_{\rm f} \simeq 1.9$.
When the dust aggregates grow into cm-sized cluster, BCCA clusters are dynamically compressed by dust-dust collisions \cite[e.g.,][]{Suyama+2008} and/or statically compressed by ram pressure of the disk gas \cite[e.g.,][]{Kataoka+2013b}.
Although it depends on the physical properties of the disk and the monomers, the compression mechanism for icy aggregates composed of submicron-size monomers in the minimum mass solar nebula \cite{Weidenschilling1977,Hayashi1981} is the static compression by ram pressure \cite{Kataoka+2013b}.
In this study, we focus on the geometrical structure of statically compressed BCCA clusters.

\begin{figure}
\centering
\includegraphics[width=4.0in]{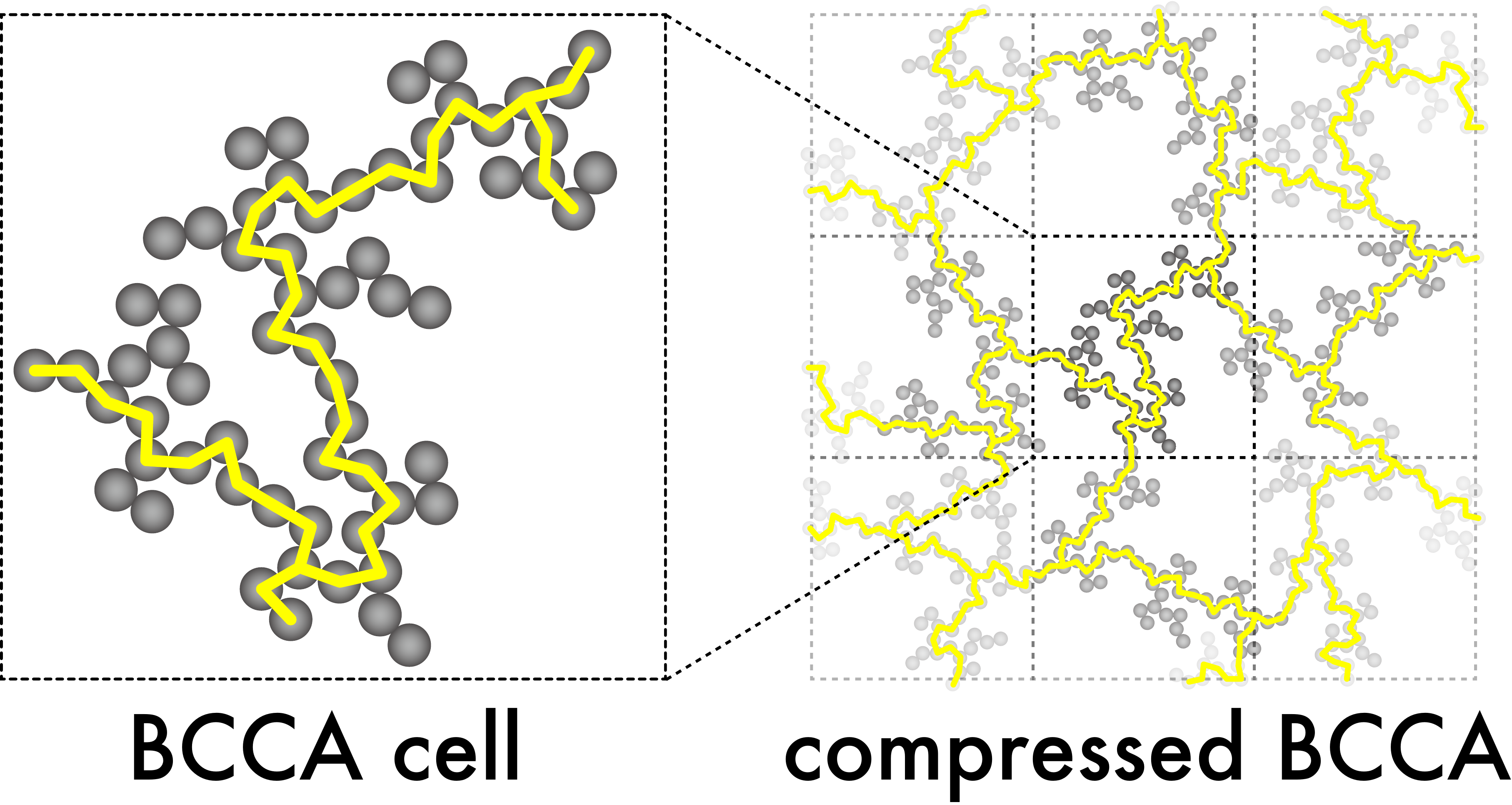}%
\caption{
Schematic description of a BCCA cell in the compressed BCCA cluster.
The yellow lines represent the possible heat paths.
}
\label{fig4}
\end{figure}

The geometrical structure of statically compressed BCCA clusters is characterized by bifractality \cite{Kataoka+2013a}.
Kataoka et al.\ \cite{Kataoka+2013a} calculated the average number of particles in spheres of radii $r_{\rm in}$, $N_{\rm in}$.
For statically compressed BCCA clusters, $N_{\rm in}$ is approximately given by
\begin{alignat}{2}
N_{\rm in} \sim {\left( \frac{r_{\rm in}}{r_{0}} \right)}^{D_{\rm f}} &\quad &&{( r_{\rm in} \ll r_{\rm tr} )},\\
N_{\rm in} \sim \phi {\left( \frac{r_{\rm in}}{r_{0}} \right)}^{3}    &      &&{( r_{\rm in} \gg r_{\rm tr} )},
\end{alignat}
and the transition radius $r_{\rm tr}$ is evaluated as $r_{\rm tr} \sim \phi^{1 / {( D_{\rm f} - 3 )}} r_{0}$.
In other words, the fractal dimension becomes three on a large scale, while it remains 1.9, i.e., $D_{\rm f}$ of BCCA, on a small scale.
This structure evolution suggests that the static compression reconstructs the fractal aggregate first on a large scale, because of the weak compressive strength on a large scale \cite{Kataoka+2013a}.
Therefore, we can imagine that it is possible to understand the physical properties of statically compressed BCCA clusters from the geometrical structure of small BCCA clusters preserved in compressed aggregates (hereinafter referred to as ``BCCA cells'').
Figure \ref{fig4} shows the schematic description of a BCCA cell in the compressed BCCA cluster.

It is important to note that the geometrical structure of dynamically compressed BCCA clusters is also characterized by bifractality \cite[e.g.,][]{Wada+2008}.
The resulting fractal dimension is approximately 2.5 on a large scale and it remains $D_{\rm f}$ of BCCA on a small scale.
Therefore, bifractality is a common characteristic of compressed BCCA clusters.
We also hypothesize that this bifractality is a common structure for compressed fractal aggregates although the initial cluster is not originated from BCCA but other aggregation processes, for example, diffusion-limited cluster aggregation \cite[e.g.,][]{Meakin1983} or reaction-limited cluster aggregation \cite[e.g.,][]{Jullien+1984}.
We will, however, need to confirm this hypothesis in the future.

\section{Thermal conductivity}
\label{sec3}

In this section, we calculate the thermal conductivity of compressed BCCA clusters and demonstrate the manner in which the geometrical structure affects the thermal conductivity.

\subsection{Methods}

We calculate the thermal conductivity of compressed BCCA clusters composed of 16384 ($= 2^{14}$) monomers.
The snapshots used in this study and used in our previous study (Arakawa et al.\ \cite{Arakawa+2019}) are the same and were prepared by Tatsuuma et al.\ \cite{Tatsuuma+2019}.
The methods of the thermal conductivity calculation are described in our previous studies \cite{Arakawa+2017,Arakawa+2019}, which we briefly summarize it here.

Dust aggregates are statically compressed in a cubic periodic boundary (see Figure 1 of Arakawa et al. \cite{Arakawa+2017}).
We consider one-directional heat flow from the lower to the upper boundary plane.
The thermal conductivity of a dust aggregate in a cubic periodic boundary $k$ is given by
\begin{equation}
k = 2 k_{\rm mat} \frac{r_{\rm c}}{r_{0}} f,
\end{equation}
where $f$ is a dimensionless function of $\phi$, $k_{\rm mat}$ is the material thermal conductivity, and $r_{\rm c}$ is the contact radius of monomer grains.
The normalized thermal conductivity $f$ is given by
\begin{equation}
f \equiv \frac{r_{0} L}{S} \sum_{\rm upper} \frac{T_{j} - T_{i}}{{\Delta}T},
\end{equation}
where $L$ is the length of the side of the cube, $S = L^{2}$ is the area of the upper and lower boundaries, $T_{i}$ is the temperature of $i$-th monomer, and ${\Delta}T$ is the temperature difference between the upper and lower boundaries.
We took the sum of contacts between the $i$-th grain on the upper boundary and $j$-th grain inside the boundaries (see Arakawa et al.\ \cite{Arakawa+2017} for details).

\begin{figure}
\centering
\includegraphics[width=2.5in]{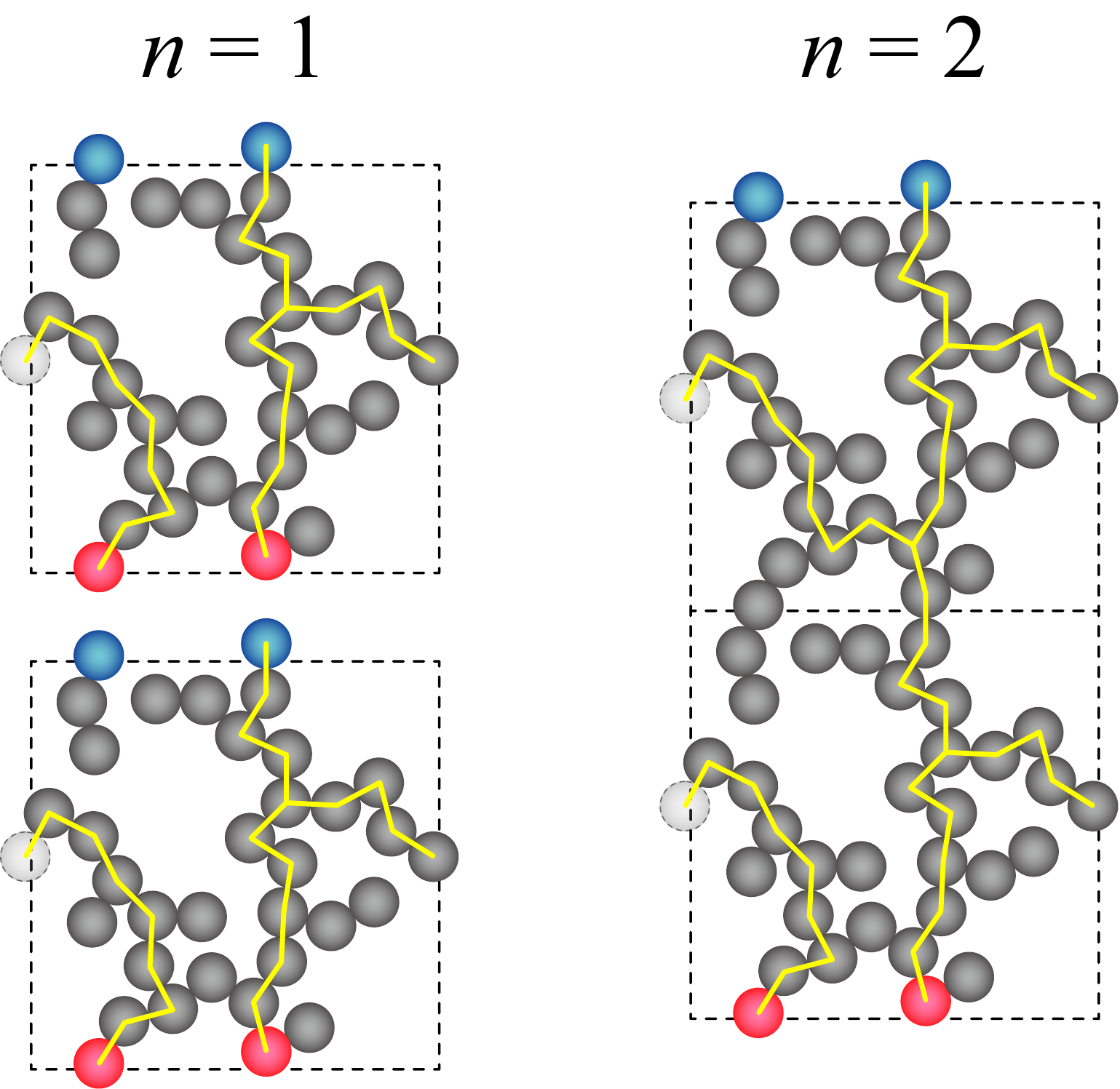}%
\caption{
Schematic description of the series connection of dust aggregates in a cubic periodic boundary.
The blue monomers are on the upper boundary and the reds are on the lower boundary, and the yellow lines represent the heat paths.
The series connection of dust aggregates would reduce the artificial effects of the boundary condition.
}
\label{fig5}
\end{figure}

In this study, we also consider the series connection of dust aggregates in a cubic periodic boundary (see Figure \ref{fig5}).
It is expected that the series connection of dust aggregates would reduce the artificial effects of the boundary condition on the thermal conductivity calculations.

\subsection{Filling factor dependence}

\begin{figure}
\centering
\includegraphics[width=3.0in]{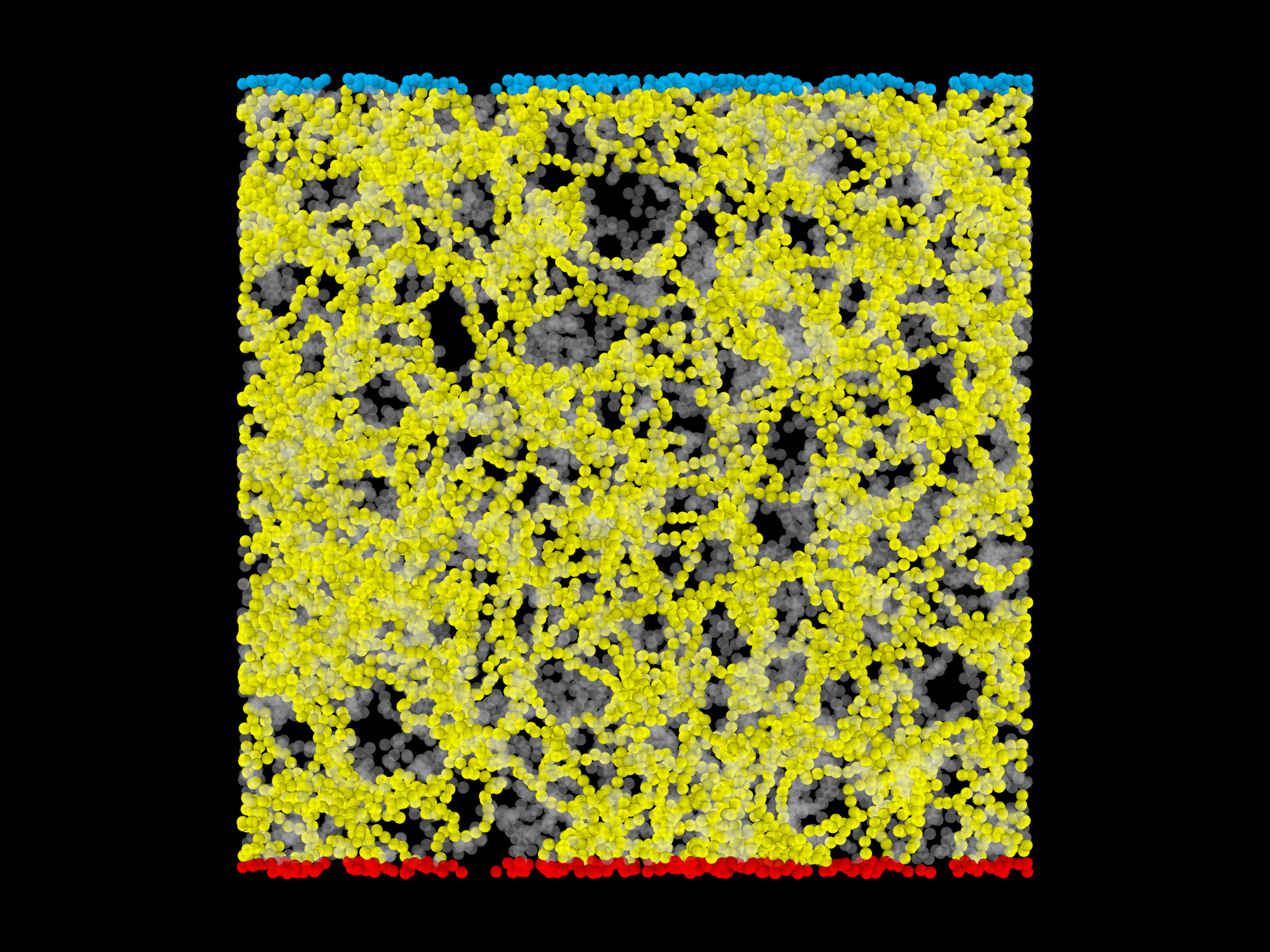}%
\caption{
An example snapshot of a compressed BCCA cluster in a cubic periodic boundary.
The blue monomers are on the upper boundary and the reds are on the lower boundary.
The yellow monomers represent the heat paths and the grays are the non-contributing monomers.
The filling factor of the aggregates is $\phi = 10^{-1.5}$.
}
\label{fig6}
\end{figure}

Figure \ref{fig6} shows the projection of three-dimensional compressed BCCA cluster in a cubic periodic boundary.
The blue monomers are on the upper boundary and the reds are on the lower boundary.
The yellow monomers represent the heat paths and the grays are the non-contributing monomers.
The filling factor of the aggregates is $\phi = 10^{-1.5}$.
It is clear that not all monomers contribute to the heat transfer within the dust aggregate.

\begin{figure}
\centering
\includegraphics[width=3.0in]{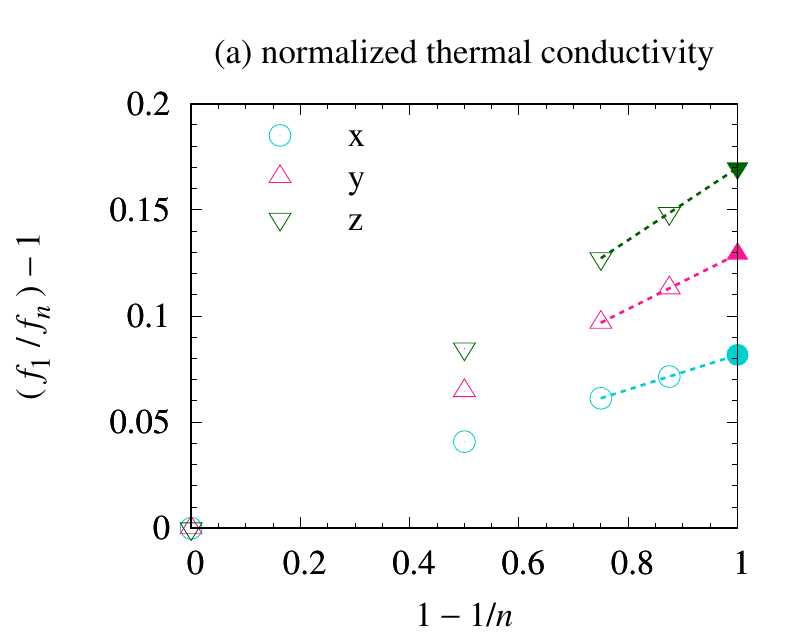}%
\includegraphics[width=3.0in]{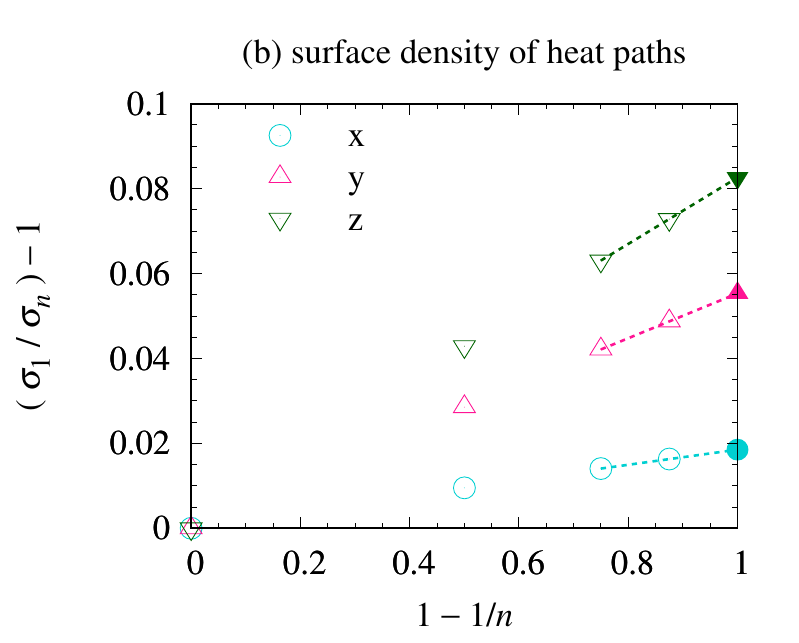}%
\caption{
(a) Example calculation of $f_{\infty}$ (filled marker) using $f_{1}$, $f_{2}$, $f_{4}$, and $f_{8}$ (open markers).
(b) Example calculation of $\sigma_{\infty}$ (filled marker) by using $\sigma_{1}$, $\sigma_{2}$, $\sigma_{4}$, and $\sigma_{8}$ (open markers).
The aggregate used in this calculation is the same that used in Figure \ref{fig6}.
We calculated $f_{\infty}$ and $\sigma_{\infty}$ from three directions ($x$, $y$, and $z$).
}
\label{fig7}
\end{figure}

It is predicted that the normalized thermal conductivity of dust aggregates in a series connection of $n$ cubes, $f_{n}$, is given by
\begin{equation}
\frac{1}{f_{n}} \simeq \frac{1}{n} {\left( \frac{1}{f_{1}} + \frac{n - 1}{f_{\infty}} \right)},
\label{eqinfty}
\end{equation}
where $f_{\infty}$ is defined as
\begin{equation}
f_{\infty} = \lim_{n \to \infty} f_{n}.
\end{equation}
We can rewrite Equation (\ref{eqinfty}) as
\begin{equation}
\frac{f_{1}}{f_{n}} - 1 \simeq {\left( \frac{f_{1}}{f_{\infty}} - 1 \right)} {\left( 1 - \frac{1}{n} \right)},
\label{eq16}
\end{equation}
and we found that ${( f_{1} / f_{n} )} - 1$ is approximately proportional to $1 - {( 1 / n )}$.
In Figure \ref{fig7}(a), we confirmed that the relation between ${( f_{1} / f_{n} )} - 1$ and $1 - {( 1 / n )}$ works well.
Therefore, we can evaluate $f_{\infty}$ using $f_{4}$ and $f_{8}$ as follows:
\begin{equation}
f_{\infty} \simeq {\left( \frac{2}{f_{8}} - \frac{1}{f_{4}} \right)}^{-1}.
\end{equation}

\begin{figure}
\centering
\includegraphics[width=3.0in]{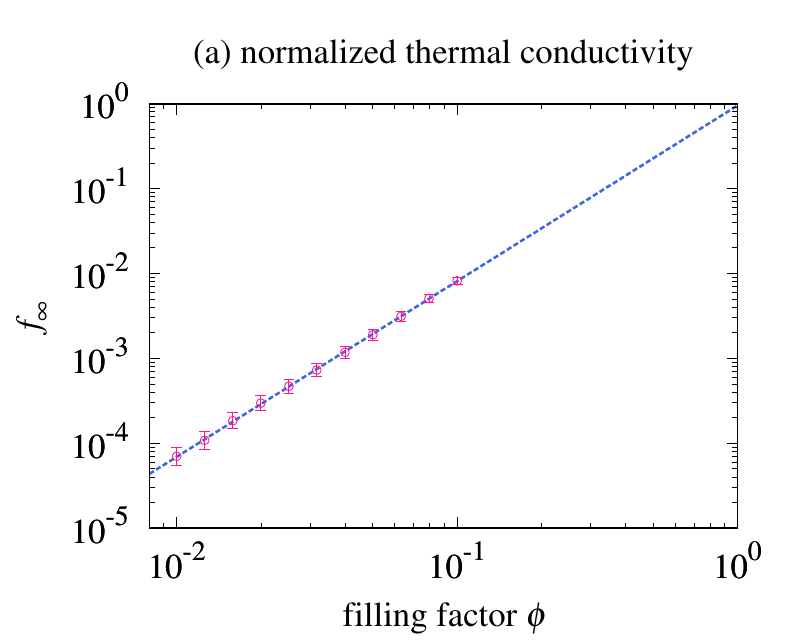}%
\includegraphics[width=3.0in]{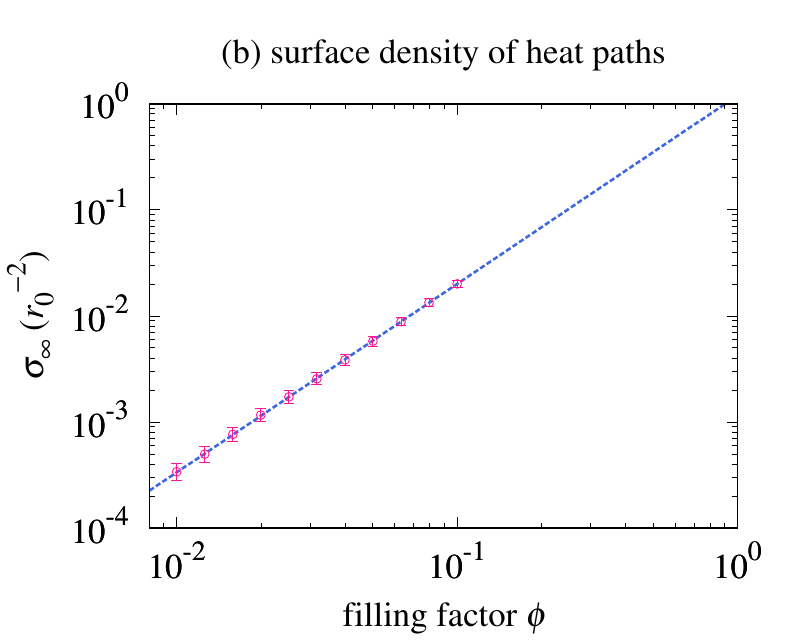}%
\caption{
(a) Fitting of the normalized thermal conductivity for the limiting case of $n \to \infty$, $f_{\infty}$, as a function of the filling factor $\phi$.
(b) Fitting of the surface density of heat paths for the limiting case of $n \to \infty$, $\sigma_{\infty}$, as a function of $\phi$.
The circles represent the averaged data with vertical error bars of twice the standard error.
The dashed line is the best-fit obtained from the weighted least-squares method.
}
\label{fig8}
\end{figure}

Figure \ref{fig8}(a) shows the normalized thermal conductivity for the limiting case of $n \to \infty$, $f_{\infty}$, as a function of the filling factor $\phi$.
We used 10 snapshot data for each $\phi$ obtained from different compression simulations \cite{Tatsuuma+2019} and calculated $f_{\infty}$ from three directions.
The circles represent the geometric mean of 30 calculation results of the temperature structure, with vertical error bars of twice the standard error.
We found that $f_{\infty}$ is given by
\begin{equation}
\log_{10} f_{\infty} = {\left( 2.068 \pm 0.034 \right)} \log_{10} \phi + {\left( -0.022 \pm 0.007 \right)},
\end{equation}
and given uncertainties are the standard errors.
This result is consistent with those of previous studies \cite{Arakawa+2017,Arakawa+2019}.

\subsection{Surface density of heat paths}

The thermal conductivity of dust aggregates must be affected by the typical length of the chain of monomers $R_{\rm geo}$ and the surface density of heat paths $\sigma$.
Here, we introduce the number of heat paths at the temperature $T$, $\mathcal{N}_{\rm path} {( T )}$.
We define $\mathcal{N}_{\rm path} {( T )}$ as the number of contacts between two monomers whose temperatures are $T_{i}$ and $T_{j}$ with $T_{i} < T < T_{j}$.
Thereafter, the average number of heat paths $\overline{\mathcal{N}_{\rm path}}$ is given by
\begin{equation}
\overline{\mathcal{N}_{\rm path}} \equiv \frac{1}{{\Delta}T} \int_{- {\Delta}T / 2}^{+ {\Delta}T / 2} {\rm d}T {\mathcal N}_{\rm path} {( T )},
\end{equation}
where the temperature at the upper and lower boundaries are $- {\Delta}T / 2$ and $+ {\Delta}T / 2$, respectively.
The surface density of heat paths $\sigma$ is given by
\begin{equation}
\sigma \equiv \frac{\overline{\mathcal{N}_{\rm path}}}{S}.
\end{equation}

The average number of heat paths $\overline{\mathcal{N}_{\rm path}}$ depends on the number of connected cubes, $n$.
In Figure \ref{fig5}, $\overline{\mathcal{N}_{\rm path}} = 3/2$ for the case of $n = 1$ and $\overline{\mathcal{N}_{\rm path}} = 409/281$ for the case of $n = 2$.
As well as $f_{\infty}$, we evaluated the surface density of heat paths for the limiting case of $n \to \infty$, $\sigma_{\infty}$.
In Figure \ref{fig7}(b), we confirmed that $\sigma_{\infty}$ can be predicted as follows:
\begin{equation}
\sigma_{\infty} \simeq {\left( \frac{2}{\sigma_{8}} - \frac{1}{\sigma_{4}} \right)}^{-1},
\end{equation}
where $\sigma_{n}$ is the surface density of heat paths within a dust aggregate in a series connection of $n$ cubes.
Figure \ref{fig8}(b) shows the surface density of heat paths for the limiting case of $n \to \infty$, $\sigma_{\infty}$, as a function of the filling factor $\phi$.
We found that
\begin{equation}
\log_{10} \frac{\sigma_{\infty}}{{r_{0}}^{-2}} = {\left( 1.775 \pm 0.025 \right)} \log_{10} \phi + {\left( 0.076 \pm 0.005 \right)},
\end{equation}
and given uncertainties are the standard errors.

For a BCCA cell, we can imagine that the average number of heat paths is
\begin{equation}
\overline{\mathcal{N}_{\rm path}} \sim \mathcal{O} {\left( 1 \right)}.
\end{equation}
Thereafter, the surface density of heat paths within a BCCA cell is given by
\begin{equation}
\sigma = \frac{\overline{\mathcal{N}_{\rm path}}}{S} \sim {R_{\rm gyr}}^{-2}.
\end{equation}
The relation between the gyration radius $R_{\rm gyr}$ and the filling factor $\phi$ is
\begin{equation}
\frac{R_{\rm gyr}}{r_{0}} \sim N^{1 / D_{\rm f}} \sim \phi^{1 / {\left( D_{\rm f} - 3 \right)}},
\end{equation}
and we obtain the relation between $\sigma$ and $\phi$:
\begin{equation}
\sigma \sim \phi^{2 / {\left( 3 - D_{\rm f} \right)}} {r_{0}}^{-2}.
\end{equation}
We found that $1.85 < D_{\rm f} < 1.92$ in Section \ref{sec2.1}, therefore, we obtain $1.74 < {2 / {( 3 - D_{\rm f} )}} < 1.85$.
The range of $2 / {( 3 - D_{\rm f} )}$ matches the numerical result, $\sigma \sim \phi^{1.775 \pm 0.025} {r_{0}}^{-2}$.
This fact validates the assumption that $\overline{\mathcal{N}_{\rm path}} \sim \mathcal{O} {( 1 )}$ for BCCA cells.

We note that the tensile strength of compressed BCCA clusters, $P_{\rm t}$, is approximately given by $P_{\rm t} \sim \sigma F_{\rm c}$, where $F_{\rm c} = {3 \pi \gamma r_{0}} / 2$ is the maximum force required to separate two sticking monomers and $\gamma$ is the surface energy \cite{Dominik+1997}.
Therefore, the tensile strength is given by
\begin{equation}
P_{\rm t} \sim \gamma {r_{0}}^{-1} \phi^{1.775 \pm 0.025},
\end{equation}
which is consistent with the numerical result of Tatsuuma et al.\ \cite{Tatsuuma+2019}, $P_{\rm t} \simeq 0.6 \gamma {r_{0}}^{-1} \phi^{1.8}$.
The coincidence of the filling factor dependence may indicate not only the number of heat paths but the number of force chains is also on the order of unity for BCCA cells.

\subsection{Understanding the filling factor dependence of the thermal conductivity}

Here, we demonstrate the manner in which the filling factor dependence of the thermal conductivity is derived from the geometrical structure.
For compressed BCCA clusters, the fractal dimension is three on a large scale, then the thermal conductivity of compressed BCCA clusters should be the same as the thermal conductivity of BCCA cells.

The spatial scale of BCCA cells is $L \sim R_{\rm gyr}$ and the area of the BCCA cells is $S \sim {R_{\rm gyr}}^{2}$, where $R_{\rm gyr} \sim N^{1 / D_{\rm f}} r_{0}$ is the gyration radius of a BCCA cell and $N$ is the number of monomers in a BCCA cell.
The surface density of heat paths is approximately given by $\sigma \sim {R_{\rm gyr}}^{-2}$.
The typical temperature difference between two contacting monomers, ${\delta}T$, is also given by
\begin{equation}
{\delta}T \sim \frac{{\Delta}T}{R_{\rm geo} / r_{0}},
\end{equation}
where ${\Delta}T$ is the temperature difference between the upper and lower region of a BCCA cell.
The heat conductance at the contact of two monomers, $H$, is \cite[e.g.,][]{Luikov+1968,Cooper+1969}
\begin{equation}
H = 2 k_{\rm mat} r_{\rm c},
\end{equation}
and the heat flow at the contact of two monomers, $I$, is $I \sim H {\delta}T$.
Therefore, the heat flow density within the BCCA cell is
\begin{equation}
k \frac{{\Delta}T}{L} \sim \sigma H {\delta}T,
\end{equation}
and the thermal conductivity $k$ is rewritten as follows:
\begin{equation}
k \sim 2 k_{\rm mat} \frac{r_{\rm c}}{r_{0}} \frac{{r_{0}}^{2}}{R_{\rm gyr} R_{\rm geo}}.
\end{equation}
The normalized thermal conductivity $f$ of the BCCA cell (and the compressed BCCA cluster) is therefore given by
\begin{equation}
f \sim \frac{{r_{0}}^{2}}{R_{\rm gyr} R_{\rm geo}} \sim N^{- {\left( 1 + \alpha \right)} / {D_{\rm f}}}.
\end{equation}
The relation between $N$ and $\phi$ is $N \sim \phi^{D_{\rm f} / {( D_{\rm f} - 3 )}}$, then we obtain the following equation:
\begin{equation}
f \sim \phi^{{\left( 1 + \alpha \right)} / {\left( 3 - D_{\rm f} \right)}} \sim \phi^{2.09}.
\end{equation}
The derived relation shows excellent coincidence with our numerical result, $f \sim \phi^{2.068 \pm 0.034}$.

\section{Discussion}
\label{sec4}

\subsection{Reinterpretation of the filling factor dependence of the compressive strength}

We can also derive the filling factor dependence of the compressive strength of compressed BCCA clusters from the geometrical structure.
In this section, we evaluate the compressive strength $P_{\rm c}$ as Kataoka et al.\ \cite{Kataoka+2013a} did.

The compressive force on the surface area of the BCCA cell $F_{\rm c}$ and the compressive strength $P_{\rm c}$ are given by
\begin{equation}
F_{\rm c} \sim P_{\rm c} {R_{\rm gyr}}^{2}.
\end{equation}
The length of the force chain within the BCCA cell is $R_{\rm geo}$.
Since the compression is accompanied by the rolling of pairs of monomers in the force chain, the work required for compression can be given by
\begin{equation}
F_{\rm c} R_{\rm geo} \sim E_{\rm roll},
\end{equation}
where $E_{\rm roll} = 6 \pi^{2} \gamma r_{0} \xi_{\rm cr}$ is the energy needed to rotate a monomer around its connection point by $\pi / 2\ {\rm rad}$ called the rolling energy, and $\xi_{\rm cr}$ is the critical rolling displacement \cite{Dominik+1997}.
Subsequently, we found that the compressive strength $P_{\rm c}$ is given by
\begin{equation}
P_{\rm c} \sim \frac{E_{\rm roll}}{{R_{\rm gyr}}^{2} R_{\rm geo}} \sim \frac{E_{\rm roll}}{{r_{0}}^{3}} \phi^{{\left( 2 + \alpha \right)} / {\left( 3 - D_{\rm f} \right)}},
\label{eqcomp}
\end{equation}
and ${\left( 2 + \alpha \right)} / {\left( 3 - D_{\rm f} \right)} \simeq 2.99$ for compressed BCCA clusters.
The derived relation shows an excellent agreement with the numerical results of Kataoka et al.\ \cite{Kataoka+2013a}, i.e.,  $P_{\rm c} \sim {( E_{\rm roll} / {{r_{0}}^{3}} )} \phi^{3}$.

We note that the original explanation by Kataoka et al.\ \cite{Kataoka+2013a} might not be accurate.
Kataoka et al.\ \cite{Kataoka+2013a} evaluated the work required for compression as
\begin{equation}
F_{\rm c} R_{\rm gyr} \sim E_{\rm roll},
\end{equation}
and the filling factor dependence of the compressive strength was obtained as
\begin{equation}
P_{\rm c} \sim \frac{E_{\rm roll}}{{R_{\rm gyr}}^{3}} \sim \frac{E_{\rm roll}}{{r_{0}}^{3}} \phi^{3 / {\left( 3 - D_{\rm f} \right)}}.
\end{equation}
This estimate was based on the assumption that the compression is accompanied by the rolling of single pair of monomers in a BCCA cell.
In this derivation, $3 / {\left( 3 - D_{\rm f} \right)} \simeq 2.69$ and it might not reproduce their numerical results.
Although our findings suggest that the $\alpha$ parameter associated with the chain length plays a significant role on the compression of dust aggregates, further studies on the force distribution within compressed fractal aggregates are required.

\subsection{Revisiting the average coordination number of compressed aggregates}

The average coordination number (i.e., the average number of contacts per monomer) $Z$ increases as an aggregate is compressed.
Arakawa et al.\ \cite{Arakawa+2019} found that, for compressed BCCA clusters, the filling factor dependence of $Z$ is given by $Z = 2 + 9.38 \phi^{1.62}$.
Here, we derive this equation from the geometrical structure.

Considering the graph structure of BCCA cluster, which is a tree (i.e., a connected acyclic graph), the average coordination number of a compressed BCCA clusters is
\begin{equation}
Z \sim \frac{2 N + C}{N},
\end{equation}
where the constant $C$ is the number of the inter-cell contacts per BCCA cell.
The number of the faces, edges, and corners within a cube is 6, 12, and 8, respectively.
Subsequently, we assume that the number of the inter-cell contacts per BCCA cell is $C \sim 9$.

We define the deviation of the coordination number from two, $\zeta \equiv Z - 2$.
The deviation $\zeta$ is given by
\begin{equation}
\zeta \sim \frac{C}{N} \sim C \phi^{D_{\rm f} / {\left( 3 - D_{\rm f} \right)}},
\end{equation}
and $1.60 < D_{\rm f} / {\left( 3 - D_{\rm f} \right)} < 1.79$ when we take the uncertainty of $D_{\rm f}$ into consideration.
Therefore, we obtain the following equation:
\begin{equation}
Z = 2 + C \phi^{D_{\rm f} / {\left( 3 - D_{\rm f} \right)}},
\end{equation}
which is consistent with the numerical result of Arakawa et al.\ \cite{Arakawa+2019}, although the uncertainty of $D_{\rm f} / {\left( 3 - D_{\rm f} \right)}$ is large and future studies on both the fractal dimension analysis and the average coordination number are essential.
We note that the fractal dimension $D_{\rm f}$ depends on the formation process of dust aggregates.
Thereofre, the average coordination number $Z$ also depends on the formation process of dust aggregates, as reported in Seizinger and Kley \cite{Seizinger+2013}.

The compressive strength $P_{\rm c}$ is also affected by the average coordination number $Z$.
If the average coordination number is $Z \simeq 2$, nearly all the monomers can roll when they are compressed.
Therefore, the interparticle force is close to the rolling friction force and the compressive strength is given by Equation (\ref{eqcomp}).
On the other hand, in the high-density region ($\phi \gg 0.1$ and $Z \gg 2$), most of the particles cannot roll freely and the compressive strength is larger than the evaluated value for the case of $Z \simeq 2$ \cite[e.g.,][]{Guettler+2009,Seizinger+2012,Omura+2017,Omura+2018}.
Then, we expect that the compressive strength would be given by the sliding friction force in the high-density limit \cite[][]{Omura+2017}, although future studies are required to understand this in detail.

\section{Summary}

In this study, we conducted the numerical simulations of the BCCA of small dust particles and calculated the geometrical structure of the fractal dust aggregates.
Additionally, we derived the filling factor dependence of the physical properties of porous dust aggregates.
Our key findings are summarized as follows.

\begin{enumerate}
\item{
We calculated the gyration radius $R_{\rm gyr}$ and the graph-based geodesic radius $R_{\rm geo}$ as the functions of the number of constituent particles $N$.
We found that $R_{\rm gyr} / r_{0} \sim N^{0.531 \pm 0.011}$ and $R_{\rm geo} / r_{0} \sim N^{0.710 \pm 0.013}$, where $r_{0}$ is the radius of constituent particles.
Thereafter, we defined two constants which characterize the geometrical structure of fractal aggregates: $D_{\rm f}$ and $\alpha$.
The definition of $D_{\rm f}$ and $\alpha$ are $N \sim {( R_{\rm gyr} / r_{0} )}^{D_{\rm f}}$ and ${R_{\rm geo}} / {r_{0}} \sim {\left( {R_{\rm gyr}} / {r_{0}} \right)}^{\alpha}$, respectively.
We revealed that $D_{\rm f} \simeq 1.88$ and $\alpha \simeq 1.34$ for BCCA clusters.
}
\item{
Kataoka et al.\ \cite{Kataoka+2013a} found that the geometrical structure of statically compressed BCCA clusters is characterized by bifractality.
This structure evolution suggests that the static compression reconstructs the fractal aggregate first on a large scale because of the weak compressive strength on a large scale.
Therefore, we can imagine that it is possible to understand the physical properties of statically compressed BCCA clusters from the geometrical structure of small BCCA clusters preserved in compressed aggregates (``BCCA cells'').
}
\item{
We investigated the filling factor dependence of thermal conductivity of statically compressed aggregates.
We found that the filling factor dependence can be interpreted from the geometrical structure of dust aggregates.
The thermal conductivity of statically compressed aggregates $k$ is given by $k \sim 2 k_{\rm mat} {( r_{\rm c} / r_{0} )} \phi^{(1 + \alpha) / (3 - D_{\rm f})}$, where $k_{\rm mat}$ is the material thermal conductivity, $r_{\rm c}$ is the contact radius of constituent particles, and $\phi$ is the filling factor of dust aggregates.
}
\item{
The compressive strength $P_{\rm c}$ is also derived from the geometrical structure as $P_{\rm c} \sim {( E_{\rm roll} / {r_{0}}^{3} )} \phi^{(2 + \alpha) / (3 - D_{\rm f})}$, where $E_{\rm roll}$ is the energy needed to rotate a monomer around its connection point by $\pi/ 2$ rad.
Our finding suggests that the $\alpha$ parameter associated with the chain length plays a significant role in the compression of dust aggregates.
In addition, the average coordination number $Z$ is derived from the geometrical structure as $Z = 2 + C \phi^{D_{\rm f} / (3 - D_{\rm f})}$, where $C \sim 9$ is the number of the inter-cell contacts per BCCA cell.
}
\end{enumerate}

\section*{Ackowledgment}

We are grateful to Misako Tatsuuma, Akimasa Kataoka, and Hidekazu Tanaka for providing snapshots of compressed BCCA clusters and the simulation code for preparing BCCA clusters.
We also thank Satoshi Okuzumi, Hiroaki Katsuragi, and Sin-iti Sirono for their fruitful discussions and comments.
S.A.\ is supported by the Grant-in-Aid for JSPS Research Fellow (JP17J06861).
This work is supported by JSPS KAKENHI grant (JP18K03721).


\begin{thebibliography}{50}

\bibitem{Sorensen2001}
C.\ M.\ {Sorensen}, {Aerosol Sci.\ Tech.}, \textbf{35}, 648 (2001)

\bibitem{Ohno+2018}
K.\ {Ohno} and S.\ {Okuzumi}, {Astrophys.\ J.}, \textbf{859}, 34 (2018)

\bibitem{Ohno+2019}
K.\ {Ohno}, S.\ {Okuzumi}, and R.\ {Tazaki}, {arXiv:1908.02201 [astro-ph.EP]}

\bibitem{Adachi+1976}
I.\ {Adachi}, C.\ {Hayashi}, and K.\ {Nakazawa}, {Prog.\ Theor.\ Phys.}, \textbf{56}, 1756 (1976)

\bibitem{Dominik+1997}
C.\ {Dominik} and A.\ G.\ G.\ M.\ {Tielens}, {Astrophys.\ J.}, \textbf{480}, 647 (1997)

\bibitem{Wada+2008}
K.\ {Wada}, H.\ {Tanaka}, T.\ {Suyama}, H.\ {Kimura}, and T.\ {Yamamoto}, {Astrophys.\ J.}, \textbf{677}, 1296 (2008)

\bibitem{Tanaka+2012}
H.\ {Tanaka}, K.\ {Wada}, T.\ {Suyama}, and S.\ {Okuzumi}, {Prog.\ Theor.\ Phys.\ Suppl.}, \textbf{195}, 101 (2012)

\bibitem{Okuzumi+2012}
S.\ {Okuzumi}, H.\ {Tanaka}, H.\ {Kobayashi}, and K.\ {Wada}, {Astrophys.\ J.}, \textbf{752}, 106 (2012)

\bibitem{Arakawa+2016}
S.\ {Arakawa} and T.\ {Nakamoto}, {Astrophys.\ J.\ Lett.}, \textbf{832}, L19 (2016)

\bibitem{Tsukamoto+2017}
Y.\ {Tsukamoto}, S.\ {Okuzumi}, and A.\ {Kataoka}, {Astrophys.\ J.}, \textbf{838}, 151 (2017)

\bibitem{Tatsuuma+2018}
M.\ {Tatsuuma}, S.\ {Michikoshi}, and E.\ {Kokubo}, {Astrophys.\ J.}, \textbf{855}, 57 (2018)

\bibitem{Meakin1988}
P.\ {Meakin}, {Adv.\ Colloid Interface Sci.}, \textbf{28}, 249 (1988)

\bibitem{Meakin1991}
P.\ {Meakin}, {Rev.\ Geophys.}, \textbf{29}, 3 (1991)

\bibitem{Blum+2008}
J.\ {Blum} and G.\ {Wurm}, {Annu.\ Rev.\ Astron.\ Astrophys.}, \textbf{46}, 21 (2008)

\bibitem{Krause+2011}
M.\ {Krause}, J.\ {Blum}, Y.\ V.\ {Skorov}, and M.\ {Trieloff}, {Icarus}, \textbf{214}, 286 (2011)

\bibitem{Sakatani+2017}
N.\ {Sakatani}, K.\ {Ogawa}, Y.-i.\ {Iijima}, M.\ {Arakawa}, R.\ {Honda}, and S.\ {Tanaka}, {AIP Adv.}, 7, 015310 (2017)

\bibitem{Kobayashi+2013}
H.\ {Kobayashi}, H.\ {Kimura}, and S.\ {Yamamoto}, {Astron.\ Astrophys.}, \textbf{550}, A72 (2013)

\bibitem{Arakawa+2017}
S.\ {Arakawa}, H.\ {Tanaka}, A.\ {Kataoka}, and T.\ {Nakamoto}, {Astron.\ Astrophys.}, \textbf{608}, L7 (2017)

\bibitem{Arakawa+2019}
S.\ {Arakawa}, M.\ {Tatsuuma}, N.\ {Sakatani}, and T.\ {Nakamoto}, {Icarus}, \textbf{324}, 8 (2019)

\bibitem{Tatsuuma+2019}
M.\ {Tatsuuma}, A.\ {Kataoka}, and H.\ {Tanaka}, {Astrophys.\ J.}, \textbf{874}, 159 (2019)

\bibitem{Kataoka+2013a}
A.\ {Kataoka}, H.\ {Tanaka}, S.\ {Okuzumi}, and K.\ {Wada}, {Astron.\ Astrophys.}, \textbf{554}, A4 (2013)

\bibitem{Wurm+1998}
G.\ {Wurm} and J.\ {Blum}, {Icarus}, \textbf{132}, 125 (1998)

\bibitem{Kempf+1999}
S.\ {Kempf}, S.\ {Pfalzner}, and Th.\ K.\ {Henning}, {Icarus}, \textbf{141}, 388 (1999)

\bibitem{Okuzumi+2009}
S.\ {Okuzumi}, H.\ {Tanaka}, and M.-a.\ {Sakagami}, {Astrophys.\ J.}, \textbf{707}, 1247 (2009)

\bibitem{Mukai+1992}
T.\ {Mukai}, H.\ {Ishimoto}, T.\ {Kozasa}, J.\ {Blum}, and J.\ M.\ {Greenberg}, {Astron.\ Astrophys.}, \textbf{262}, 315 (1992)

\bibitem{Tazaki+2016}
R.\ {Tazaki}, H.\ {Tanaka}, S.\ {Okuzumi}, A.\ {Kataoka}, and H.\ {Nomura}, {Astrophys.\ J.}, \textbf{823}, 70 (2016)

\bibitem{Liu+1995}
C.-h.\ {Liu}, S.\ R.\ {Nagel}, D.\ A.\ {Schecter}, S.\ N.\ {Coppersmith},\ S.\ {Majumdar}, O.\ {Narayan}, and T.\ A.\ {Witten}, {Science}, \textbf{269}, 513 (1995)

\bibitem{Sirono2014}
S.-i.\ {Sirono}, {Meteor.\ Planet.\ Sci.}, \textbf{49}, 109 (2014)

\bibitem{Suyama+2008}
T.\ {Suyama}, K.\ {Wada}, and H.\ {Tanaka}, {Astrophys.\ J.}, \textbf{684}, 1310 (2008)

\bibitem{Kataoka+2013b}
A.\ {Kataoka}, H.\ {Tanaka}, S.\ {Okuzumi}, and K.\ {Wada}, {Astron.\ Astrophys.}, \textbf{557}, L4 (2013)

\bibitem{Weidenschilling1977}
S.\ J.\ {Weidenschilling}, {Astrophys.\ Space Sci.}, \textbf{51}, 153 (1977)

\bibitem{Hayashi1981}
C.\ {Hayashi}, {Prog.\ Theor.\ Phys.\ Suppl.}, \textbf{70}, 35 (1981)

\bibitem{Meakin1983}
P.\ {Meakin}, {Phys.\ Rev.\ Lett.}, \textbf{51}, 1119 (1983)

\bibitem{Jullien+1984}
R.\ {Jullien} and M.\ {Kolb}, {J.\ Phys.\ A}, \textbf{17}, L639 (1984)

\bibitem{Luikov+1968}
A.\ V.\ {Luikov}, A.\ G.\ {Shashkov}, L.\ L.\ {Vasiliev}, and Yu.\ E.\ {Fraiman}, {Int.\ J.\ Heat Mass Transfer}, \textbf{11}, 117 (1968)

\bibitem{Cooper+1969}
M.\ G.\ {Cooper}, B.\ B.\ {Mikic}, and M.\ M.\ {Yovanovich}, {Int.\ J.\ Heat Mass Transfer}, \textbf{12}, 279 (1969)

\bibitem{Seizinger+2013}
A.\ {Seizinger} and W.\ {Kley}, {Astron.\ Astrophys.}, \textbf{551}, A65 (2013)

\bibitem{Guettler+2009}
C.\ {G\"{u}ttler}, M.\ {Krause}, R.\ J.\ {Geretshauser}, R.\ {Speith}, and J.\ {Blum}, {Astrophys.\ J.}, \textbf{701}, 130 (2009)

\bibitem{Seizinger+2012}
A.\ {Seizinger}, R.\ {Speith}, and W.\ {Kley}, {Astron.\ Astrophys.}, \textbf{541}, A59 (2012)

\bibitem{Omura+2017}
T.\ {Omura} and A.\ M.\ {Nakamura}, {Planet.\ Space Sci.}, \textbf{149}, 14 (2017)

\bibitem{Omura+2018}
T.\ {Omura} and A.\ M.\ {Nakamura}, {Astrophys.\ J.}, \textbf{860}, 123 (2018)


\end{thebibliography}
\end{document}